\documentclass[conference]{IEEEtran}
\IEEEoverridecommandlockouts

\usepackage{cite}
\usepackage{amsmath,amssymb,amsfonts}
\usepackage{algorithmic}
\usepackage{graphicx}
\usepackage[dvipsnames,table]{xcolor}
\usepackage[final]{microtype}
\usepackage[italic]{mathastext}
\usepackage{libertine}
\usepackage[T1]{fontenc}
\usepackage{textcomp}
\usepackage[varqu,varl]{zi4}
\usepackage[all]{nowidow}
\usepackage[keeplastbox]{flushend}
\usepackage{fancyhdr}
\usepackage{stmaryrd}
\usepackage{float}

\usepackage{xspace}
\usepackage{subcaption}
\usepackage{colortbl}
\usepackage{multirow}
\usepackage{booktabs}
\usepackage{listings}

\usepackage{tikz}

\newcommand{\minisection}[1]{\vspace{1ex}\noindent {\bf #1}}

\usepackage{lipsum}

\definecolor{heatred}{RGB}{255, 102, 102}
\definecolor{heatorange}{RGB}{255, 178, 102}
\definecolor{heatyellow}{RGB}{255, 255, 153}
\definecolor{heatgreen}{RGB}{204, 255, 204}

\usepackage[linecolor=red,backgroundcolor=red!25,bordercolor=red,textsize=tiny]{todonotes}

\usepackage{hyperref}
\usepackage{fontawesome}
\usepackage[most]{tcolorbox}
\usepackage{etoolbox} 
\newcounter{insightcounter}

\newcounter{questioncounter}
\newcommand{\question}[2][]{
  \refstepcounter{questioncounter}
  \begin{tcolorbox}[
    colframe=gray,
    colback=white,
    boxrule=0.5pt,
    arc=3pt,
    left=0pt,
    right=0pt,
    top=0pt,
    bottom=0pt
  ]
  \ifx\\#1\\%
    \textbf{\faQuestionCircle\ Q\arabic{questioncounter}}: #2
  \else
    \hypertarget{#1}{}%
    \label{#1}%
    \textbf{\faQuestionCircle\ \hyperref[a:#1]{Q\ref*{#1}}}: #2
  \fi
  \end{tcolorbox}
}

\newcommand{\sym}[1]{\textsf{#1}}
\newcommand{\name}{\sym{CryptOracle}\xspace}
\newcommand{\thetitle}{\name: A Modular Framework to Characterize FHE}

\definecolor{termblue}{RGB}{40,70,200}
\definecolor{termgreen}{RGB}{0,140,0}

\lstdefinestyle{terminal}{
  basicstyle=\ttfamily\small,
  columns=fullflexible,
  keepspaces=true,
  showstringspaces=false,
  breaklines=true,
  escapeinside={(*@}{@*)},
}

\newtcblisting{terminalbox}{
  enhanced,
  breakable,
  listing only,
    listing options={
    style=terminal,
    mathescape=false,
    texcl=false,
    escapechar=,
    escapeinside={},
  },
  colback=white,
  colframe=black,
  boxrule=0.6pt,
  arc=0pt,
  left=6pt,
  right=6pt,
  top=6pt,
  bottom=6pt,
  boxsep=0pt,
  width=\columnwidth
}

\begin{document}


\title{\thetitle}

\author{\IEEEauthorblockN{Cory Brynds}
\IEEEauthorblockA{\textit{ECE Department} \\
\textit{University of Central Florida}\\
Orlando, FL \\
cory.brynds@ucf.edu}
\and
\IEEEauthorblockN{Parker McLeod$^*$\thanks{$^*$This work was completed while at the University of Central Florida.}}
\IEEEauthorblockA{\textit{Advanced Micro Devices} \\
Folsom, CA \\
pmcleod@amd.com}
\and
\IEEEauthorblockN{Lauren Caccamise$^*$}
\IEEEauthorblockA{\textit{ECE Department} \\
\textit{Purdue University}\\
West Lafayette, Indiana \\
lcaccami@purdue.edu}
\and
\IEEEauthorblockN{Asmita Pal$^\dagger$\thanks{$^\dagger$This work was completed while at the University of Wisconsin-Madison.}}
\IEEEauthorblockA{\textit{Advanced Micro Devices} \\
Santa Clara, CA \\
asmitpal@amd.com}
\and
\IEEEauthorblockN{Dewan Saiham}
\IEEEauthorblockA{\textit{CREOL} \\
\textit{University of Central Florida}\\
Orlando, FL \\
dewan.saiham@ucf.edu}
\and
\IEEEauthorblockN{Sazadur Rahman}
\IEEEauthorblockA{\textit{ECE Department} \\
\textit{University of Central Florida}\\
Orlando, FL \\
mohammad.rahman@ucf.edu}
\and
\IEEEauthorblockN{Joshua San Miguel}
\IEEEauthorblockA{\textit{ECE Department} \\
\textit{University of Wisconsin-Madison}\\
Madison, WI \\
jsanmiguel@wisc.edu}
\and
\IEEEauthorblockN{Di Wu}
\IEEEauthorblockA{\textit{ECE Department} \\
\textit{University of Central Florida}\\
Orlando, FL \\
di.wu@ucf.edu}
}

\maketitle



\begin{abstract}
Privacy-preserving machine learning has become an important long-term pursuit in this era of artificial intelligence (AI). 
Fully Homomorphic Encryption (FHE) is a uniquely promising solution, offering provable privacy and security guarantees. 
Unfortunately, computational cost is impeding its mass adoption. 
Modern solutions are up to six orders of magnitude slower than plaintext execution. 
Understanding and reducing this overhead is essential to the advancement of FHE, particularly as the underlying algorithms evolve rapidly.
This paper presents a detailed characterization of OpenFHE, a comprehensive open-source library for FHE, with a particular focus on the CKKS scheme due to its significant potential for AI and machine learning applications. We introduce \name, a modular evaluation framework comprising (1) a benchmark suite, (2) a hardware profiler, and (3) a predictive performance model. The benchmark suite encompasses OpenFHE kernels at three abstraction levels: workloads, microbenchmarks, and primitives. The profiler is compatible with standard and user-specified security parameters. \name monitors application performance, captures microarchitectural events, and logs power and energy usage for AMD and Intel systems. 
These metrics are consumed by a modeling engine to estimate runtime and energy efficiency across different configuration scenarios, with prediction error ranging from $-7.02\%\sim8.40\%$ for runtime and $-9.74\%\sim15.67\%$ for energy (geomean).
\name is open source, fully modular, and serves as a shared platform to facilitate the collaborative advancements of applications, algorithms, software, and hardware in FHE.
The \name code can be accessed at https://github.com/UnaryLab/CryptOracle.
\end{abstract}
\section{Introduction}
\label{sec:Introduction}

Modern computation is increasingly offloaded to the cloud, where shared infrastructure and limited transparency heighten the risk of data compromise.  
Contemporary security frameworks routinely encrypt data in storage and during transmission; however, they ultimately require plaintext access for computation. 
This window of exposure allows adversaries to exploit memory access patterns~\cite{memoryTraceOblivious,camouflage}, leverage microarchitectural side channels~\cite{svf,osvikcache2005}, or trigger latent bugs in complex software stacks~\cite{foreshadow,heartbleed,spectre}, especially under the proliferation of machine learning based attacks~\cite{MembershipAttack,GANAttack}.

Fully Homomorphic Encryption (FHE) offers a compelling alternative by enabling computation on encrypted data without decryption~\cite{gentry2009fully}. 
However, this guarantee comes with a much higher cost as compared to other secure-computation paradigms, not just because of the computational intensity but also due to the evolving nature of workloads in the FHE space.
Consequently, the systems community has pursued aggressive optimizations to overcome these challenges,
catalyzing a vibrant ecosystem of architectural and algorithmic innovations, spanning general-purpose GPU implementations~\cite{fan2023tensorfhe} to highly specialized hardware accelerators~\cite{samardzic2021f1, samardzic2022craterlake, kim2022bts, kim2022ark, agrawal2023fab}.

\begin{figure}
    \centering
    \includegraphics[width=\columnwidth]{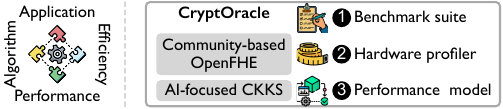}
    \caption{Overview of this paper.
    \label{fig:high_level}
    }
\end{figure}

\minisection{Problem Statement.}
Though existing profilers~\cite{pal2023fully,jiang2022fhebench,hebench} provide valuable insights, a gap remains in \emph{automating rapid characterization across abstraction levels, hardware vendors, and evaluation metrics}. 
We highlight challenges as follows.

\begin{enumerate}
    \item \textbf{Long runtime.} FHE operations are intrinsically expensive due to the reliance on modular polynomial arithmetic and number-theoretic transforms, e.g., a multiplication takes 0.156ms in FHE vs. 4.3ns in plaintext~\cite{9481143}.
    This makes simulation unsuitable, typically adopted in the architectural space, as it would be orders of magnitude slower.
    \item \textbf{Complicated optimization space.} Performance in FHE is influenced by interdependent factors such as ciphertext size, multiplicative depth, noise growth, parallelism and memory layout. 
    For example, ciphertexts can exceed 50~MB~\cite{kim2022bts} and increase memory pressure; bootstrapping, taking up to 1.5s~\cite{samardzic2021f1}, is needed for noise management.
    Hence navigating this space involves tuning multiple architectural and algorithmic parameters, making design-space exploration less intuitive.
    \item \textbf{Rapid algorithmic change.} Due to the strong guarantees of secure computation, the FHE system is evolving rapidly. Frequent updates to encoding strategies and security standards make it difficult to build reusable performance models.
\end{enumerate}

\minisection{The Solution.}
We introduce \name, a modular framework for systematic characterization of FHE applications, as shown in Figure~\ref{fig:high_level}. 
Built atop the community-maintained \textsc{OpenFHE} library~\cite{OpenFHE}, \name lowers the barrier to reproducible evaluation for theorists, system architects, and application developers alike.  It comprises:

\begin{enumerate}
    \item \textbf{Benchmark suite.}  A top-down collection of representative ML workloads, optimized microbenchmarks, and the semantic primitives (semantically atomic, like multiplication) that underlie them.  
    \item \textbf{Hardware profiler.}  An automated tool that extracts runtime, power, and microarchitectural statistics on both AMD and Intel CPUs, averaged across multiple runs under varied parameter sets.
    \item \textbf{Performance model.}  A lightweight additive model that extrapolates primitive-level measurements to predict end-to-end runtime and energy within milliseconds, enabling efficient design-space exploration.
\end{enumerate}

\minisection{Contributions.}
\begin{itemize}
    \item We present \name, the first open-source, end-to-end characterization framework for CKKS FHE workloads on commodity CPUs.
    \item We curate and release a comprehensive benchmark suite spanning popular ML applications, targeted microbenchmarks, and underlying primitives.
    \item We develop an automated profiler that collects unified performance, power, and security metrics across processor vendors and parameterized inputs.
    \item We devise a fast prediction model that estimates performance from primitive-level statistics.
    \item We validate the model experimentally, distill architectural insights, and illustrate how \name accelerates research on FHE algorithms and systems.
\end{itemize}

The remainder of the paper is organized as follows.
Section~\ref{sec:Background} presents the motivation behind this work. Section~\ref{sec:Architecture} introduces the components of the \name framework, followed by implementation details in Section~\ref{sec:Implementation}. Section~\ref{sec:Evaluation} provides an empirical evaluation of the system. We discuss limitations in Section~\ref{sec:Limitation}, review prior works in Section~\ref{sec:relatedWork} and conclude in Section~\ref{sec:Conclusion}.
\section{Background and Motivation}
\label{sec:Background}

\subsection{FHE Framework}

Introduced by Gentry in 2009\cite{gentry2009fully}, FHE allows computations to be performed directly on encrypted data, enabling privacy-preserving computation in untrusted environments. 
Following Gentry's original construction, which was initially kept theoretical due to prohibitive computational overhead, subsequent advancements have led to the development of multiple FHE schemes that improve efficiency and functionality. 
Notable schemes include BGV\cite{brakerski2014leveled}, BFV\cite{fan2012somewhat}, FHEW\cite{ducas2015fhew}, TFHE\cite{chillotti2020tfhe}, and CKKS\cite{cheon2017homomorphic}.

The increasing interest in deploying FHE in real-world applications has catalyzed the development of several FHE libraries, including Microsoft SEAL\cite{sealcrypto}, PALISADE\cite{polyakov2017palisade}, HEAAN\cite{cheon2017homomorphic}, Lattigo\cite{lattigo}, HELayers\cite{aharoni2020helayers} and OpenFHE\cite{OpenFHE}, each supporting a certain number of different schemes.
Among these, OpenFHE is one of the most comprehensive libraries. 
First, OpenFHE supports a wide range of FHE schemes under a unified API. 
Second, it offers compatibility with the Homomorphic Encryption Standard\cite{lauter2022protecting}, ensuring cryptographic soundness and reproducibility. 
Third, OpenFHE allows fine-grained customization of encryption parameters, enabling performance-security tradeoff studies across varied hardware backends. 
Finally, it features built-in support for both leveled and bootstrapped CKKS, facilitating benchmarking across arbitrary depths.
OpenFHE supports a rich set of homomorphic operations, including ciphertext-plaintext and ciphertext-ciphertext addition and multiplication, as well as advanced primitives such as vector rotation, scalar encoding, conjugation, bootstrapping.

\subsection{The CKKS Scheme}

\begin{figure}[!t]
  \centering
  \includegraphics[width=0.5\textwidth]{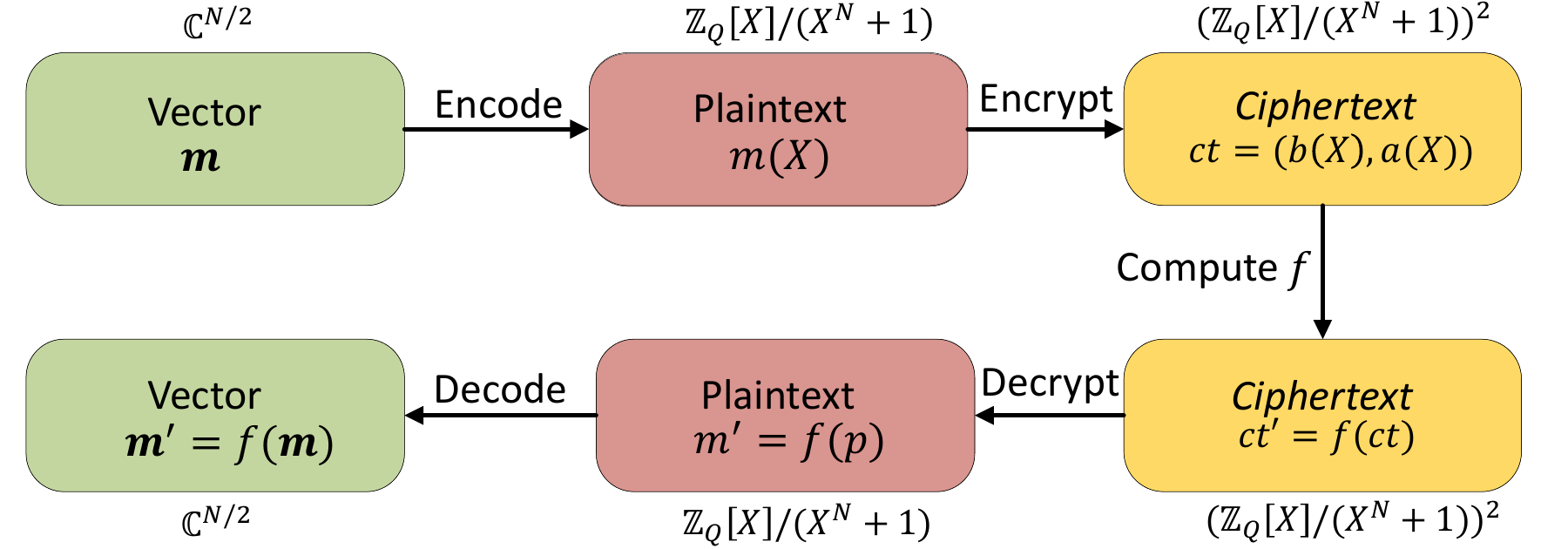}
  \caption{Overview of CKKS scheme.
  \label{fig:overview_ckks}
  }
\end{figure}

The CKKS scheme supports approximate floating-point arithmetic over vectors of complex numbers, making it particularly suited for applications in machine learning, analytics, and scientific computing. 
Figure~\ref{fig:overview_ckks} illustrates the process. A complex plaintext vector ${m} \in \mathbb{C}^{N/2}$ is first encoded as the polynomial $m(X) = \sum_{i=0}^{N-1} c_i X^i$ in the ring $R_Q = \mathbb{Z}_Q[X]/(X^N + 1)$, where $N$ is a power of two and $Q$ is the ciphertext modulus. Encryption samples a random polynomial $a(X)$ and a small Gaussian error $e(X)$, then outputs the ciphertext $\text{ct} = (b(X), a(X)) = (a(X) \cdot s(X) + m(X) + e(X),\ a(X))$, where $s(X)$ is the secret key. Decryption recovers the message by computing $m'(X) = ct \cdot  (1 - s(X) )= m(X) + e(X)$, and $m(X)$ can be obtained by removing the noise term $e(X)$.

Homomorphic operations in CKKS operate over encrypted vectors. 
With \( \llbracket \mathbf{m} \rrbracket \) referring to the ciphertext of a plaintext message of \( \mathbf{m} \), important operations include:

\begin{itemize}
    \item \textbf{Add}: Given ciphertexts $\llbracket \mathbf{m}_1 \rrbracket$ and $\llbracket \mathbf{m}_2 \rrbracket$, produce $\llbracket \mathbf{m}_1 + \mathbf{m}_2 \rrbracket$, where the addition is performed component-wise over $\mathbb{C}$.
    \item \textbf{Mult}: Given ciphertexts $\llbracket \mathbf{m}_1 \rrbracket$ and $\llbracket \mathbf{m}_2 \rrbracket$, produce $\llbracket \mathbf{m}_1 \circ \mathbf{m}_2 \rrbracket$, where $\circ$ denotes element-wise multiplication.
    \item \textbf{Rotate}: For a ciphertext $\llbracket \mathbf{m} \rrbracket$ and an integer $k$, compute $\llbracket \phi_k(\mathbf{m}) \rrbracket$, where $\phi_k$ cyclically rotates the vector by $k$ positions. For example, when $k = 1$: $\mathbf{x} = (x_0, x_1, \ldots, x_{n-2}, x_{n-1})$, and $\phi_1(\mathbf{x}) = (x_{n-1}, x_0, \ldots, x_{n-3}, x_{n-2})$.
\end{itemize}

CKKS leverages the Number Theoretic Transform (NTT) to accelerate polynomial multiplication: once mapped to the NTT domain, the operation reduces to a point-wise product, reducing the computational complexity from $\mathcal{O}(N^2)$ to $\mathcal{O}(N \log N)$.

Each homomorphic multiplication amplifies the accumulated noise and simultaneously scales the plaintext. To curb this noise growth, the ciphertext is rescaled by dividing it by its last modulus prime $q_L$~\cite{cheon2018full}: \( \text{Rescale}(\text{ct}) \rightarrow \text{ct}' = \text{ct}/q_L \mod Q' \), where \( Q' = Q/q_L \). After several multiply–rescale rounds, the remaining modulus eventually becomes too small to accommodate further multiplications, capping the achievable circuit depth.
Because CKKS does not natively support nonlinear functions (e.g. ReLU), computations rely on polynomial approximations, which typically demand a substantial multiplicative depth\cite{lee2023precise}.
To permit computations of arbitrary depth, CKKS employs \emph{bootstrapping}\cite{gentry2009fully}.
This procedure homomorphically evaluates the decryption followed by the re-encryption of a ciphertext, thereby resetting its noise budget and restoring both its modulus and available multiplicative depth.
Consequently, encrypted data can undergo an unlimited sequence of operations while still producing correct results.
Bootstrapping is indispensable, yet it remains computationally intensive and consumes several levels of the ciphertext modulus. State-of-the-art implementations accelerate bootstrapping through FFTs and polynomial approximations\cite{cheon2018bootstrapping}, making it the focus of ongoing research in hardware and algorithmic optimizations.

\subsection{Benchmarking and Characterization}

Benchmarking and workload characterization are essential for understanding application behavior and for guiding both system- and compiler-level optimizations. Over time, diverse methodologies have been devised to capture the performance characteristics of both general-purpose and domain-specific workloads. General-purpose benchmark suites such as SPEC CPU2017~\cite{spec_cpu2017}, Graph500~\cite{graph500}, and MLPerf~\cite{mattson2020mlperftrainingbenchmark} have driven advances in evaluating compute-intensive, irregular-memory-acess, and deep-learning applications. In parallel, specialized characterization frameworks have been proposed to target particular system behaviors. Bircher\cite{bircher2011complete}, for instance, leveraged hardware performance counters for comprehensive system-level power estimation, whereas Crape\cite{crape2020rigorous} developed a statistically rigorous methodology for profiling dynamic Python workloads. Liu\cite{liu2020deffe} further extended this direction by proposing a data-efficient framework tailored to domain-specific architectures. Despite their breadth, these studies concentrate on conventional, plaintext applications and therefore overlook the unique challenges of FHE, whose encrypted computation, intricate algebraic operations, and unique performance bottlenecks demand specialized profiling strategies.

\subsection{Motivation of \name}

\name is built on OpenFHE, targeting CKKS across three abstraction levels: workload, microbenchmark, and primitive.
The hardware profiler captures CPU behaviors at the primitive level and reports a diverse set of hardware metrics.
Using the runtime and energy measured for primitives, the performance model predicts the corresponding values for workloads and microbenchmarks.

\question{Why benchmark CKKS on OpenFHE?}

Recent advances in FHE software and hardware acceleration have moved FHE closer to practical deployment~\cite{ebel2025orion, park2025fhendi, agrawal2023fab, samardzic2022craterlake}. Even with this progress, the community still lacks a unified, systematic framework for benchmarking. 

We adopt OpenFHE because it is community-driven, open source, actively maintained, and executable on widely available CPU platforms. This community-driven backbone of \name on AMD and Intel processors lets a broad audience study application and algorithm characteristics under identical conditions, promoting fair comparison and accelerating research progress.

Although OpenFHE also implements BGV and BFV, our study focuses on CKKS, the scheme that supports floating-point arithmetic central to AI workloads, to keep workload semantics, runtime behavior, and metric interpretation consistent. Future work may extend the methodology to multiple schemes.

\question{Why profile FHE at the primitive level?}

FHE introduces computational patterns that differ fundamentally from those in conventional benchmarks such as intensive polynomial arithmetic, explicit ciphertext management, and intricate noise control mechanisms.

Where suites such as SPEC or MLPerf evaluate general-purpose programs, FHE workloads intertwine cryptographic kernels with high-dimensional algebraic routines and operations that map poorly to commodity hardware.
Consequently, \name profiles execution at the primitive level. 
This approach, common in HPC and accelerator design, leverages the consistent and predictable behavior of primitives to guide optimizations across all layers.
Profiling with primitive granularity exposes system bottlenecks in fine detail, enabling targeted optimization~\cite{jung2021over, fan2023tensorfhe}.

Finally, the core CKKS primitives (MULT, RESCALE, ADD, and ROTATE) are expected to remain stable. Future applications will compose these operations rather than redefine them, so \name provides enduring value for performance analysis.

\question{Why model and predict CPU performance?}

The high computational cost of FHE has motivated extensive work on custom hardware accelerators\cite{samardzic2021f1, samardzic2022craterlake, kim2022bts, kim2022ark, agrawal2023fab}. These efforts usually appear alongside new applications and algorithms. Because implementing each idea on actual hardware is complex and time-consuming, assessing the benefits only after deployment is impractical. A faster approach is to estimate performance and efficiency on general-purpose CPUs through analytical or empirical models. Similar modeling has become standard practice for predicting the runtime of AI workloads at various scales\cite{qi17paleo, calculon_paper,10.1145/3669940.3707265}. Following this methodology, \name estimates the performance and energy of FHE applications from primitive-level profiling, enabling accurate yet rapid exploration of algorithmic and application innovations.

\section{\name Framework}
\label{sec:Architecture}

\begin{figure*}[!t]
    \centering
    \includegraphics[width=\textwidth]{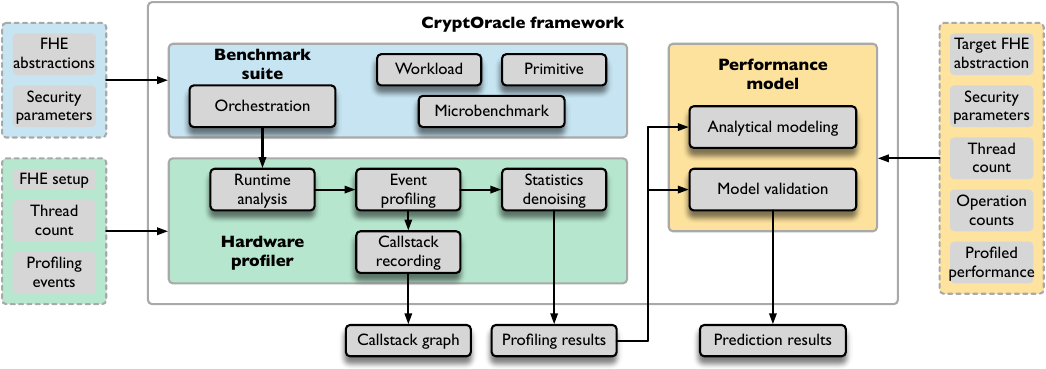}
    \caption{Overview of \name framework.}
    \label{fig:overview_benchmark}
\end{figure*}

\subsection{Overview}
Our \name framework characterizes the performance of FHE through a comprehensive benchmark suite and an integrated hardware profiler. In addition, its performance model enables \name to rapidly evaluate the performance of the application as the security and hardware parameters are varied.
Our \name framework is shown in Figure~\ref{fig:overview_benchmark}, with inputs and outputs listed in Table~\ref{tab:framework_io}.

\begin{table}[!t]
\centering
\caption{\name framework I/O overview.
\label{tab:framework_io}
}
\begin{tabular}{l|l|l}
\toprule
\multicolumn{2}{l|}{\textbf{I/O Type}} & \textbf{Variable} \\
\midrule
\midrule
\multirow{9}{*}{Input}
  & Benchmark & \textbf{FHE abstractions}: FHE workloads,\\
  & suite & microbenchmarks, primitives \\
  & & \textbf{Security parameters}: ciphertext ring \\
  & & dimension, security standard, batch size,  \\
  & & computation depth \\
  \cmidrule{2-3}
  & Hardware & \textbf{FHE setup}: run count for FHE \\
  & profiler & abstractions, a list of abstractions to profile \\
  & & \textbf{Thread count}: max allowable OMP threads\\
  & & \textbf{Profiling events}: a list of hardware events \\
  & &  to record during profiling \\
  \cmidrule{2-3}
  & Performance & \textbf{Target abstraction}: microbenchmark \\
  & model & or workload\\
  & & \textbf{Security parameters}: OpenFHE \\
  & & cryptocontext parameters\\
  & & \textbf{Thread count}: number of OMP threads\\
  & & \textbf{Operation counts}: FHE operation counts \\
  & & from the workload and microbenchmark \\
  & & \textbf{Profiled performance}: runtime, \\
  & & energy of profiled FHE abstractions  \\
\midrule
\multicolumn{2}{l|}{Intermediate} & Profiling raw recordings \\
\midrule
\multicolumn{2}{l|}{Output} & Callstack graph, profiling/prediction results   \\
\bottomrule
\end{tabular}
\end{table}

\subsection{Benchmark Suite}
\label{sec:benchmark}
The \name benchmark suite is the primary interface for end users of the framework.  
It packages a predefined set of FHE abstractions: primitives, microbenchmarks, and workloads, each ready for profiling.  
After the user supplies the desired security parameters, the suite instructs the orchestrator to perform a combinatorial sweep over the parameters of all abstractions.  
When multiple abstraction layers are selected, the orchestrator evaluates them in ascending order of complexity: primitives first, followed by microbenchmarks, and finally workloads.  
For every parameter combination, the orchestrator passes the relevant configuration to the hardware profiler. Table~\ref{tab:crypto-config} summarizes all FHE abstractions currently supported.

\begin{table}[htbp]
\centering
\caption{Supported microbenchmarks and workloads in the \name benchmark suite, with the default application parameters used for evaluation.
\label{tab:crypto-config}}
\begin{tabular}{l|l|l|l|l|l}
\toprule
\textbf{Category} & \textbf{Application} & \textbf{Sec. Std.} & \textbf{N} & \textbf{B} & \textbf{L} \\
\midrule
\midrule
Primitive& OpenFHE Primitives & none & $2^{16}$ & $2^{12}$ & 10 \\
\midrule
\multirow{3}{*}{Micro-}&Matrix Multiplication & none & $2^{16}$ & $2^{12}$ & 10 \\
&Logistic Function & none & $2^{16}$ & $2^{12}$ & 10 \\
benchmark&Sign Evaluation & none & $2^{16}$ & $2^{12}$ & 10 \\
\midrule
\multirow{4}{*}{Workload} &CIFAR-10 Inference & none & $2^{16}$ & $2^{12}$ & 5 \\
&Chi Square Test & none & $2^{17}$ & 1 & 3 \\
&ResNet20 Inference & none & $2^{16}$ & $2^{14}$ & 10 \\
&Logreg Training & 128-bit & $2^{17}$ & $2^{16}$ & 14\\
\bottomrule
\end{tabular}
\end{table}

\subsection{Hardware Profiler}
\label{sec:profiler}
The hardware profiler executes four sequential phases. 
First, during runtime analysis it measures the application's overall execution time and energy consumption. 
Second, during event profiling it counts hardware performance events using the same whole-application and setup executions. 
Third, in the denoising phase it subtracts the setup measurements from both the runtime and event data to obtain metrics that exclude one-time I/O activity and profiling overhead. 
Finally, it records the application's call stack to support call stack graph generation.

\subsubsection{Runtime Analysis}
During the runtime analysis phase, the hardware profiler records both execution time and energy consumption of the target application with Perf, a user-space Linux utility that interfaces with CPU performance monitoring units (PMUs).
Perf estimates energy use through the Running Average Power Limit (RAPL) interface, which the CPU periodically samples integrated energy counters.
With \name, we query the RAPL Package domain, which captures the cumulative power drawn by the entire processor socket (cores, caches, and attached memory).

Runtime metrics are collected in this same pass because their acquisition introduces negligible overhead compared with event profiling or call-stack tracing.
For each primitive operation, the profiler repeats the measurement several times and reports the median to reduce variance, a practice that is especially important for low-latency primitives.

Runtime analysis requires two passes. The first “regular” pass measures the full workload (e.g., matrix multiplication plus ciphertext setup through deserialization). The second “setup” pass isolates the initialization cost, measuring only the runtime and energy consumption of the setup steps (i.e., the initial deserialization of ciphertexts).

\subsubsection{Event Profiling}
The second phase of the hardware profiler carries out detailed event profiling with Linux Perf.
Perf samples hardware registers at a specified frequency to count microarchitectural events.
Users can choose any PMU events supported by the host CPU. Common selections include retired instructions, branches, branch mispredictions, and total cache references, as summarized in Table~\ref{tab:perf_event}.

For each unique benchmark configuration supplied by the orchestrator, the profiler executes the workload N times, where N is the user defined run count. The thread count is recorded as part of every event-profiling run.

\begin{table}[!h]
\centering
\caption{Performance events currently collected.\label{tab:perf_event}}
\begin{tabular}{l|p{6cm}}
\toprule
\textbf{Category} & \textbf{Event} \\
\midrule
\midrule
Core & instructions, cpu-cycles, branches, branch-misses \\
\midrule
Cache & cache-references, cache-misses, L1-dcache-loads, L1-icache-load-misses \\
\midrule
TLB & dTLB-loads, dTLB-load-misses, iTLB-load-misses \\
\midrule
Page Faults & page-faults, minor-faults \\
\bottomrule
\end{tabular}
\end{table}

\subsubsection{Statistics Denoising}

After profiling, the raw statistics recorded during the setup phase are subtracted from those collected during the steady-state execution phase for each timing, energy, and perf counter, isolating the program’s region of interest (ROI).
For low-latency primitives, these values are averaged across all internal calls to ensure a sufficiently large sample.
Next, average power consumption is calculated.

All metrics are stored for subsequent data analysis and performance model construction.

\subsubsection{Callstack Recording}
\begin{figure}
    \centering
    \includegraphics[width=\columnwidth]{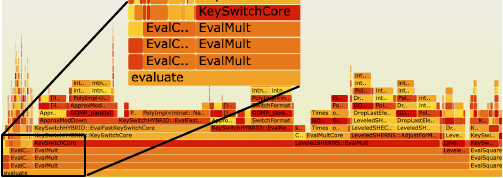}
    \caption{FlameGraph of the logistic function.
    \label{fig:example_flamegraph}}
\end{figure}

During a third pass of the hardware profiler, users can record the application's call stack.
Capturing this information reveals which FHE operations dominate execution time.
The resulting stack trace is automatically rendered as a FlameGraph~\cite{gregg_flamegraph_2017}. In a FlameGraph, the width of each box indicates the proportion of total runtime spent in the corresponding function, while the vertical axis represents call-stack depth.
Figure~\ref{fig:example_flamegraph} presents the logistic function, showing that roughly 90\% of its runtime is consumed by the homomorphic multiply operation.
This visualization enables developers to identify performance bottlenecks at early stages.

\subsection{Performance Model}
Packaged with the hardware profiler is a lightweight performance model that rapidly evaluates application runtime and other metrics for security parameters and thread counts that have not yet been tested.
We use \textit{application} to denote both microbenchmarks and full workloads.
The model receives two inputs: the primitive profiling results and the counts of every homomorphic operation in the target application.
It then linearly extrapolates the application runtime (e.g. ResNet-20 inference) by summing, for each primitive, the product of its count and its measured runtime at the chosen security parameters and thread count.
Because FHE abstractions exhibit long runtimes, this linear approach provides projections with minimal overhead.
This model could then be easily used to evaluate the impact of algorithmic changes to primitive operations or scaling an application to a higher security level.
To assess the model’s fidelity, \name includes a routine that validates the estimated runtime and reports the estimation error.

\section{Experimental Setup}
\label{sec:Implementation}

\subsection{Software Setup}

To understand the trade-offs between computational efficiency and security, we sweep the ciphertext ring dimension N across $[2^{13}, 2^{17}]$ and the computational depth L across [10, 20], maintaining a constant batch size B. We also profile the predefined security standards (128, 192, and 256-bit security from~\cite{HomomorphicEncryptionSecurityStandard}), which override the default parameter selection. To ensure a fair comparison, security parameters, evaluation keys, and input ciphertexts are generated once, serialized to files, and then reused by all primitives and microbenchmarks. 
Each workload fixes any additional application-specific parameters.
Table~\ref{tab:crypto-config} lists the default settings used in the evaluation, unless stated otherwise.
During data collection,

the median of five runs is reported for each metric. For primitive profiling, where an individual operation may complete in tens of microseconds, we invoke each primitive repeatedly until the cumulative runtime exceeds 500ms, then compute per-call averages based on the total number of invocations.

\subsection{Hardware Setup}
Profiling was conducted on two CPU platforms, an Intel i9 and an AMD Ryzen 9 (see Table~\ref{tab:hardware-spec}), to identify hardware-level performance differences between them. In OpenFHE, primitives and lower-level operations are parallelized with OpenMP. To study performance at each abstraction layer, we varied the number of OpenMP threads by setting the environment variable \texttt{OMP\_NUM\_THREADS}. Our evaluation targets fine-grained multithreading, where threads operate within a single primitive. In contrast, coarse-grained multithreading, in which threads are distributed across primitives, is disabled for both the microbenchmark and workload abstraction levels.

\begin{table}[!t]
\centering
\caption{Hardware specifications.}
\begin{tabular}{l|l|l}
\toprule
\textbf{Specification} & \textbf{Intel i9-13900K} & \textbf{AMD Ryzen 9 7950X} \\
\midrule
\midrule
Core/Thread Count & 24/32 & 16/32 \\
\midrule
CPU Freq (GHz) & 5.8 & 5.88 \\
\midrule
L1d/L1i Cache (KB) & 896/1331 & 512/512 \\
\midrule
L2/L3 Cache (MB) & 32/36 & 16/64 \\
\midrule
Total DRAM (GB) & 64 & 96 \\
\bottomrule
\end{tabular}
\label{tab:hardware-spec}
\end{table}

\section{Evaluation}
\label{sec:Evaluation}

\subsection{Profiling Results}
\subsubsection{Primitive Analysis}
\begin{figure}[!t]
  \centering
  \includegraphics[width=\linewidth]{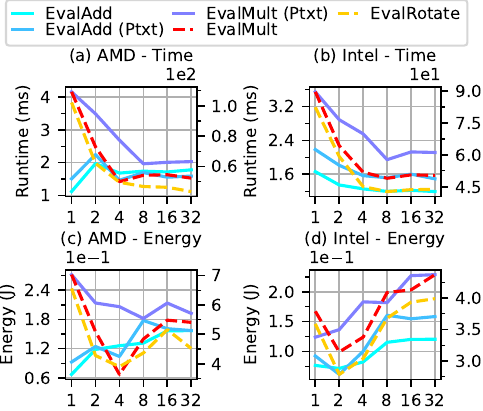}
  \caption{Runtime and energy consumption of each primitive across the evaluated CPU vendors. The horizontal axis shows the thread counts, and the vertical axis is split into a left axis (cold-colored solid lines) and a right axis (warm-colored dashed lines). Together, these curves summarize performance across all tested platforms.
  \label{fig:primitive_runtime_energy_profiled}}
\end{figure}

\begin{figure}[!t]
  \centering
  \includegraphics[width=\linewidth]{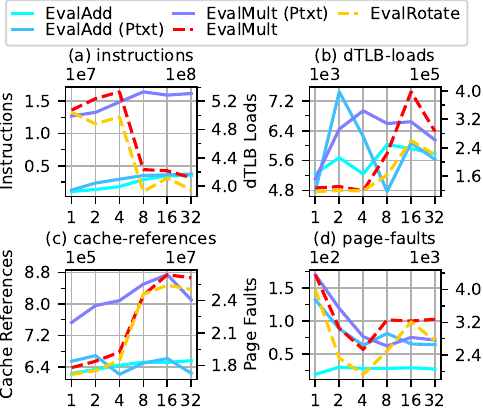}
  \caption{Primitive architecture statistics on an AMD CPU plotted against thread count on the x axis.
  \label{fig:primitive_arch_stats}}
\end{figure}

Figure~\ref{fig:primitive_runtime_energy_profiled} presents runtime and energy consumption for homomorphic encryption primitives on AMD and Intel CPUs under varying thread counts.
On both platforms, the runtime of \textit{EvalMult} and \textit{EvalRotate} falls as the number of threads rises, but the benefit plateaus at 8 threads, suggesting limited scalability and a potential memory bottleneck. Figure~\ref{fig:primitive_arch_stats}(c) supports this interpretation, showing elevated cache reference counts on AMD CPUs beyond 8 threads.
Because \textit{EvalAdd} and \textit{EvalAdd (Ptxt)} already run in about $\sim$2ms, they leave little room for further speed-up, and the OpenMP overhead incurred when moving from one to two threads can even cause a slight slowdown.

Energy consumption follows the same pattern as runtime across all kernels. Lightweight kernels maintain nearly constant energy use across thread counts, whereas compute-intensive primitives such as \textit{EvalMult} and \textit{EvalRotate} show an initial drop in energy with additional threads due to shorter runtimes, followed by an increase beyond 4 threads as parallelization overheads emerge. 
The results generated with \name emphasize its capacity to capture precise performance metrics across diverse hardware platforms.

Figure~\ref{fig:primitive_arch_stats} explores the microarchitectural behaviour of the primitives on the AMD platform, presenting a subset of the events listed in Table~\ref{tab:perf_event}. At low thread counts, \textit{EvalMult} and \textit{EvalRotate} record the largest instruction footprints, underscoring their computational intensity. As the thread count rises from one to four, the instruction counts drop substantially, most notably for \textit{EvalMult}, indicating that multithreading removes redundant serial work. In contrast, cache references increase with additional threads because of greater data sharing. The pattern of dTLB loads follows the cache trend, as increased parallel memory accesses raise TLB pressure and introduce page-level contention. Correspondingly, page-fault rates fall as more threads are added. Lightweight primitives such as \textit{EvalAdd} show only modest changes across all counters. As homomorphic encryption kernels continue to evolve, \name will offer detailed insights into their architectural challenges.

\begin{figure}[!t]
  \centering
  \includegraphics[width=\linewidth]{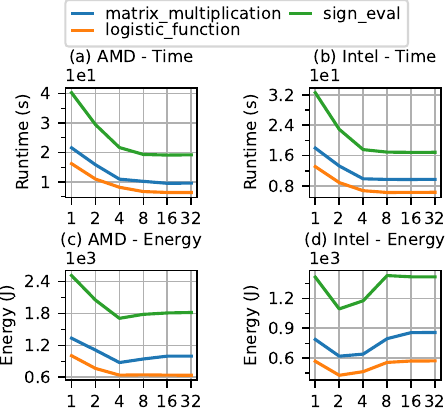}
  \caption{Runtime and energy microbenchmarks across CPU vendors with thread count plotted on the x axis.
  \label{fig:microbenchmark_runtime_energy_profiled}}
\end{figure}

\subsubsection{OpenFHE Hardware Backends}
\begin{figure}[!t]
  \centering
        \includegraphics[width=\linewidth]{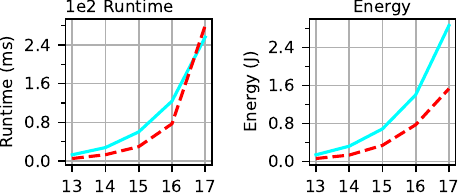}
        \caption{Single-threaded runtime and energy consumption of baseline (solid) and HEXL (dashed) backends for the EvalMult primitive across ring dimensions on AMD CPU.}
        \label{fig:hexl-comparison}
\end{figure}

While this work explores profiling OpenFHE on CPUs, OpenFHE is designed as a modular library able to interface with GPUs and accelerators via a hardware backend, if available. 
Currently, the only officially-supported OpenFHE backend is the Homomorphic Encryption Accelerator Library (HEXL). 
This library accelerates integer arithmetic on devices supporting AVX512 instructions. 
Figure~\ref{fig:hexl-comparison} compares an EvalMult operation's runtime and energy consumption with and without HEXL acceleration, which both exhibit similar trends for increasing ring dimension. 
HEXL provides an average speedup of 1.8x, with a maximum speedup of 2.36x at $N=2^{16}$, after which memory becomes the bottleneck in the operation at $N=2^{17}$.

\subsubsection{Microbenchmark Analysis}

We further analyze three microbenchmarks and several matrix sizes for matrix multiplication. A similar profile could be generated for the full workloads, but we omit it for brevity.

Figure~\ref{fig:microbenchmark_runtime_energy_profiled} presents both runtime and energy for the three microbenchmarks. As expected, increasing the number of CPU threads reduces runtime on both AMD and Intel processors. The benefit, however, diminishes quickly: the runtime curve flattens at about 8 threads on each platform.

Energy consumption follows a different pattern. The minimum energy appears at 4 threads on the AMD system and at 2 threads on the Intel system for almost every microbenchmark. Beyond these points, the small runtime savings cannot offset the additional energy required by extra threads. On the Intel CPU, using 8 to 32 threads even consumes more energy than a single thread. One likely reason is the heterogeneous design of the Intel i9-13900K, which combines high-performance (P) cores with energy-efficient (E) cores and complicates thread scheduling. These observations reveal a clear performance-efficiency trade-off that can guide the design of more sustainable applications.

\begin{figure}[!t]
  \centering
    \includegraphics[width=\linewidth]{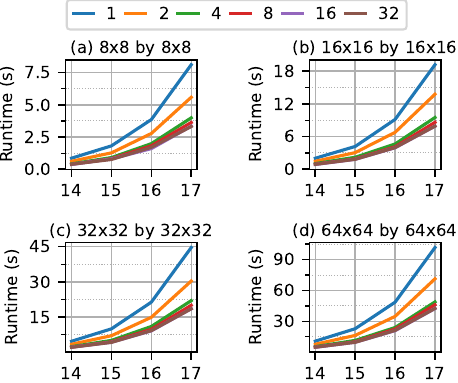}
  \caption{Matrix Mult runtime measured on an AMD CPU, with separate panels for matrix sizes, the x axis for ring dimension (powers of 2), and color coding for thread counts.}
  \label{fig:matrix_mult_runtime_profiled}
\end{figure}

Figure~\ref{fig:matrix_mult_runtime_profiled} depicts the runtime scaling of matrix multiplication.
When scaling up the ring dimension (from $2^{14}$ to $2^{17}$), the runtime grows almost linearly.
Holding the ring dimension fixed and enlarging the matrix size also yields a linear rise, for instance an approximately 2.25$\times$ slowdown when the ring dimension is 17 with one thread (the highest point on the blue curve).
This linear behavior under FHE contrasts with the quadratic scaling characteristic of plaintext matrix multiplication.
Finally, as the thread count increases, larger matrices reach saturation sooner, causing the corresponding curves to converge.

\subsubsection{Profiling Overhead}

Table~\ref{tab:profiling_overhead} summarizes the profiling overhead incurred by \name. The program region of interest encompasses only the FHE computation itself. The runtime analysis overhead corresponds to the initialization cost required to deserialize ciphertexts. This cost is relatively low and vendor independent, about 0.05s on both AMD and Intel CPUs. The event-profiling overhead reflects the extra time needed to gather performance counters. This overhead is substantially larger than that of runtime analysis and varies with both the microbenchmark and the vendor. AMD displays relatively stable overheads of roughly 60\%, whereas Intel exhibits greater fluctuation but a lower relative overhead.

\begin{table}[!t]
    \centering
    \caption{Profiling overhead of microbenchmarks over ROI for 8 threads, N=16, 
    and depth=10.
    `-A' and `-I' denote AMD and Intel CPUs.
    $\Delta\%$ denotes the relative overhead.
    \label{tab:profiling_overhead}}
    \begin{tabular}{l|c|c|c}
    \toprule
        \multirow{3}{*}{\textbf{Name}} & \multicolumn{3}{c}{\textbf{Time (s)}} \\
    \cmidrule{2-4}
         & \multirow{2}{*}{{\textbf{ROI}}} & \textbf{Runtime} & \textbf{Event} \\
         &  & \textbf{analysis ($\Delta\%$)} & \textbf{profiling ($\Delta\%$)} \\
    \midrule
        Logistic Func-A & 6.68 & 0.04 (0.60\%) & 4.06 (60.8\%) \\
        Matrix Mult-A & 10.16 & 0.04 (0.39\%) & 5.78 (56.9\%) \\
        Sign Eval-A & 19.20 & 0.05 (0.26\%) & 11.83 (61.6\%)\\
        Logistic Func-I & 6.24 & 0.05 (0.80\%) & 3.39 (54.3\%) \\
        Matrix Mult-I  & 9.77 & 0.05 (0.51\%) & 2.75 (28.1\%) \\
        Sign Eval-I & 16.78 & 0.05 (0.30\%) & 5.43 (32.4\%) \\
    \bottomrule
    \end{tabular}
\end{table}

\begin{figure}[!t]
  \centering
        \includegraphics[width=\linewidth]{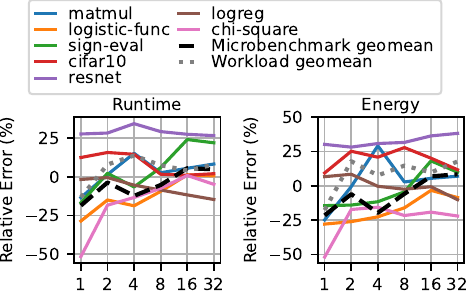}
        \caption{Prediction analysis of relative error across threads on AMD CPU.}
        \label{fig:amd_predicted_error}
\end{figure}

\subsection{Performance Model}
\label{sec:perf-model-eval}
\subsubsection{Runtime and Energy Prediction}
There are two observations in Figure~\ref{fig:amd_predicted_error}.
First, in general, the model exhibits better prediction accuracy compared to workloads, since microbenchmarks have been well optimized for FHE challenges~\cite{FhermaChallenges}, while workloads are more for functional evaluations.
Second, larger thread counts give better accuracy, as the performance saturates more.
The geomean of prediction accuracy in dashed lines concurs with the two observations above.
The overall geomean across all thread counts is -7.02\% and 8.40\% for 
microbenchmark and workload runtime, and -9.74\% and 15.67\% for microbenchmark and workload energy.

\begin{figure}[!t]
  \centering
    \includegraphics[width=\linewidth]{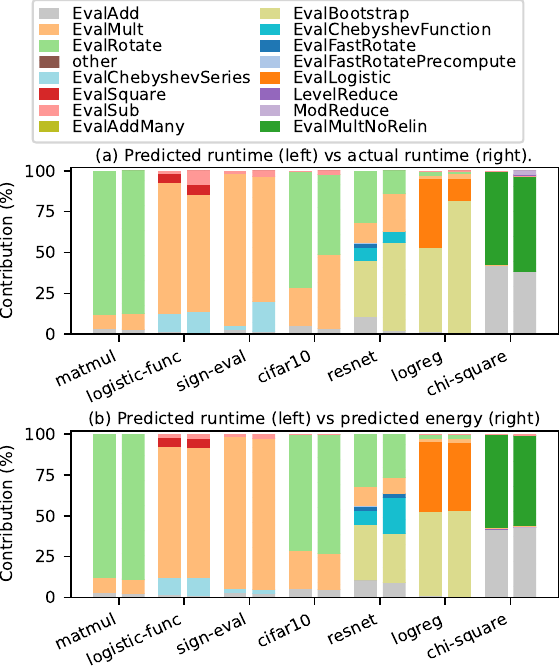}
  \caption{Primitive-level contributions for runtime and energy on an AMD CPU using 8 threads, shown across microbenchmarks and workloads (x-axis).
  \label{fig:perf_model_breakdown}}
\end{figure}

\subsubsection{Prediction Breakdown} 
Figure~\ref{fig:perf_model_breakdown} shows the contribution of primitives, expressed as a percentage of runtime or energy.
Figure~\ref{fig:perf_model_breakdown} (a) shows the predicted vs actual runtime breakdown for various workloads, while Figure~\ref{fig:perf_model_breakdown} (b) contrasts predicted runtime vs predicted energy contributions. Each bar is stacked by the type of primitive kernels used in the application. Though we show a thorough breakdown, there is scope for "other" newer kernels to be supported by \name for future workloads.
Figure~\ref{fig:perf_model_breakdown} (a) indicates strong alignment between predicted and actual runtime across microbenchmarks. While this deviates more for workloads (e.g. cifar10, resnet and logreg) due to their more complex nature, the prediction preserves the relative importance of primitives.
Figure~\ref{fig:perf_model_breakdown} (b) highlights the difference between predicted runtime and predicted energy, with most applications showing a similar distribution for each.
Resnet exhibits a slight deviation, with EvalChebyshevFunction occupying a higher energy ratio than runtime. 
Thus, our results indicate that \name accurately preserves the relative contribution of individual primitives, capturing consistent trends in both runtime and energy profiles.

\subsubsection{Prediction Overhead}
We present the prediction overhead of \name in Table~\ref{tab:prediction_overhead}.
The table compares the runtime analysis time, which measures execution time and energy without collecting performance counters, against the model prediction time, which is an almost constant cost incurred by running the Python-based performance model.
Profiling latency varies considerably across workloads, and Intel CPUs consistently outperform AMD CPUs because of their larger on-chip memory capacities, in agreement with the primitive and microbenchmark results.
The performance model accelerates runtime and energy analysis by $10\times\sim500\times$, depending on the workload.

\begin{table}[!t]
    \centering
    \caption{Prediction overhead of workloads for 8 threads.
    `-A' and `-I' denote AMD and Intel CPUs.
    The profiling time here is for runtime analysis.
    \label{tab:prediction_overhead}}
    \begin{tabular}{l|c|c|c}
    \toprule
        \multirow{2}{*}{\textbf{Name}} & \multicolumn{2}{c|}{\textbf{Time (s)}} & \multirow{2}{*}{\textbf{Speedup}}\\
    \cmidrule{2-3}
         & \textbf{Profiling} & \textbf{Prediction} & \\
    \midrule
    \midrule
        Chi Square Test-A & 49.3 & \multirow{8}{*}{0.6} & 82.2$\times$ \\
        CIFAR-10-A & 2.05 &  & 3.4$\times$ \\
        Logistic Regression-A & 253.2 & & 422.0$\times$ \\
        ResNet-20-A & 236.8 & & 394.7$\times$ \\
        
        Chi Square Test-I & 35.5 &  & 59.2$\times$ \\
        CIFAR-10-I & 1.85 &  & 3.1$\times$ \\
        Logistic Regression-I & 231.5 &  & 385.8$\times$ \\
        ResNet-20-I & 217.5 &  & 362.5$\times$ \\
    \bottomrule
    \end{tabular}
\end{table}

\subsubsection{Case Study: Algorithmic Innovation}
\begin{figure}[!t]
    \centering
    \includegraphics[width=\columnwidth]{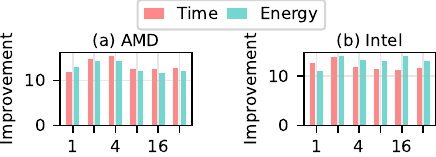}
    \caption{Estimated improvements of runtime and energy across different CPU vendors and thread counts (x axis). 
    The algorithm configurations are d=128, r=2, N=16, depth=3 for ``Algorithm 5'', and depth=6 for ``Rectangular MM''~\cite{rho2024encryption}.
    \label{fig:case_study}}
\end{figure}

To illustrate another application of \name, we estimate the runtime and energy savings that can be achieved through algorithmic innovations. Specifically, we count the primitive operations required by two algorithms implementing ciphertext–ciphertext matrix multiplication~\cite{rho2024encryption}. The projected gains are presented in Figure~\ref{fig:case_study}. Across various CPU vendors and thread counts, the projected speedups and energy reductions are consistent, ranging from $10\times\sim15\times$. The original study in~\cite{rho2024encryption} evaluated its implementation on a GPU; the measured speedup there is approximately $5.42\times$, which differs from our CPU-based projection.

\section{Limitations of This Work}
\label{sec:Limitation}

We identify two primary limitations in this work.
First, \name currently supports only CPU platforms due to the lack of supported OpenFHE hardware backends. This limitation does not diminish the merit of \name, as it remains an accessible platform for examining advances in FHE applications and algorithms. Also, GPU acceleration of FHE has shown great promise, and \name's infrastructure can be expanded to support future GPU extensions to OpenFHE. Some of these extensions already exist~\cite{agullódomingo2025fideslibfullyfledgedopensourcefhe}, but they are experimental and lack any full applications to benchmark.
Second, \name does not yet predict performance or efficiency on unseen hardware or algorithm configurations.
This gap could be addressed by machine learning-based techniques instead of simple linear extrapolation.
Similar strategies have been extensively studied for AI workloads~\cite{qi17paleo, calculon_paper, 10.1145/3669940.3707265}, and we leave their adoption to future work.


\section{Related Work}
\label{sec:relatedWork}

Although traditional benchmark suites such as 
SPEC CPU2017~\cite{spec_cpu2017}, MLPerf~\cite{mattson2020mlperftrainingbenchmark}, 
and Graph500~\cite{graph500} have been extensively studied, 
systematic performance evaluation of FHE remains sparse, but on the horizon.

Pal et al.\cite{pal2023fully} present a foundational study using OpenFHE\cite{OpenFHE}, SEAL~\cite{sealcrypto}, and HEaaN~\cite{cheon2017homomorphic}, revealing runtime and energy trade-offs across security levels, polynomial degrees, and threading models. FHEBench~\cite{jiang2022fhebench} supplies a standardized suite for latency and memory comparison across libraries, and Bergerat et al.\cite{bergerat2023parameter} explore parameter optimization for TFHE.

HEBench~\cite{hebench}, created by Intel and collaborators, evaluates canonical workloads such as dot products and matrix multiplication through a modular backend loader. Despite its robustness, it currently supports only a small number of libraries. HEProfiler~\cite{takeshita2025heprofiler} offers cycle accurate profiling of FHE primitives, exposing threading and memory inefficiencies. The native benchmark suite in OpenFHE~\cite{OpenFHE} provides only basic profiling, limited to runtime, CPU core, and frequency statistics. It omits advanced performance metrics such as latency, power, and throughput and lacks event profiling, runtime tracing, and other profiling capabilities.

These studies supply valuable empirical evidence but remain focused on microbenchmarking and parameter tuning. In contrast, \name profiles both primitives and end-to-end workloads while also delivering accurate, predictive models that guide system-level optimization.

\section{Conclusion}
\label{sec:Conclusion}
As privacy-preserving machine learning becomes increasingly critical, FHE offers strong theoretical security guarantees but also incurs considerable computational overhead, restricting practical deployment.
We introduced \name, a modular characterization framework that integrates a benchmark suite, a hardware profiler, and a performance model to understand the runtime and energy consumption of CPUs running OpenFHE.

Our open-source release of \name aims to accelerate progress toward making FHE practical for real-world machine learning workloads.

\section*{Acknowledgment}
This research is in part sponsored by the AMD University Program, in the form of machine donation and student fellowships to the University of Central Florida.

\bibliographystyle{IEEEtranS}
\bibliography{reference}


\end{document}